\documentclass[3p,preprint,11pt]{elsarticle}

\usepackage{wrapfig}
\usepackage{subfigure}
\usepackage{graphicx}
\usepackage{color}
\usepackage{balance}
\usepackage{comment, endnotes}
\usepackage{amsmath}

\usepackage{afterpage}
\usepackage{multirow}
\usepackage{float, lscape}
\usepackage{url}
\usepackage{setspace}
\usepackage[export]{adjustbox}
\usepackage{caption}

\definecolor{blue}{rgb}{0.0,0.0,1}
\newcommand*{\revise}{\textcolor{black}}

\journal{Pervasive and Mobile Computing}

\begin{document}

\begin{frontmatter}

\title{A Comprehensive Survey on Networking over TV White Spaces}

\author{Mahbubur Rahman and Abusayeed Saifullah}
\address{Department of Computer Science, Wayne State University, Detroit, MI 48202}
\ead{(r.mahbub, saifullah)@wayne.edu}

\begin{abstract}
\revise{{\em TV white spaces} refer to the allocated but locally unused TV spectrum and can be used by unlicensed 
devices as secondary users. Thanks to their lower frequencies (e.g., 54 -- 698 MHz in the US), communication over the TV spectrum has excellent propagation characteristics over long distances and through obstacles. 
These characteristics along with their wide availability make the TV white spaces a great choice and alternative to many existing wireless technologies, especially the ones that need long range and high bandwidth communication. 
In the last decade, there have been numerous efforts from academia, industries, and standards bodies for exploiting the potentials of the TV white spaces for several applications including wireless broadband Internet access. Their characteristics and features also hold potentials for many new applications including sensing and monitoring, Internet of Things, wireless control, smart utility, location-based services, and transportation and logistics.
In this paper, we perform a retrospective review and comparative study of existing work on networking in the TV white spaces. 
Additionally, we discuss the associated research challenges such as dealing with the interference between primary and secondary TV spectrum users, TV white space temporal and spatial variations and fragmentation, antenna design, mobility, and security. We also describe the future research directions to handle the above challenges. To the best of our knowledge, this is the first comprehensive survey on the literature of TV white space networking research.}
\end{abstract}


\end{frontmatter}

\section{Introduction}\label{label:intro}


In a historic ruling in 2008, the Federal Communication Commission (FCC) of the US allowed unlicensed devices (e.g., Wi-Fi devices) to operate in unoccupied TV channels. The allocated but locally unused TV channels (54 -- 698 MHz in the US) are called the {\em TV white spaces (TVWS)}. Unlicensed users such as Wi-Fi, ZigBee, and Bluetooth are allowed to access and operate in the TVWS as secondary users, i.e., they must not interfere the TV stations or other licensed users. To access TVWS, an unlicensed device can either query a cloud-hosted geo-location database or perform sensing operation to determine the energy of the spectrum~\cite{FCC_first_order}. In 2010, FCC mandated that an unlicensed device must query a database to learn about the TVWS in its location, and kept the sensing technique as an optional choice~\cite{fcc_second_order}. Similar regulations have been adopted by several other countries including UK, Canada, Singapore, Malaysia, and South Africa~\cite{wifi_like_connectivity}.

Since the TV transmissions are in the lower frequencies -- VHF and lower UHF (54 to 698 MHz) --  TVWS have excellent propagation characteristics over long distances. They can easily penetrate walls and other objects, and hence hold enormous potential for wireless applications that need long transmission range. Additionally, TVWS may be widely available in most places. Statistically, rural and suburban areas offer a larger set of TVWS compared to the urban regions due to very few official TV stations~\cite{FCC_first_order}. TVWS may also provide diverse bandwidth choices for applications with substantial bandwidth and high bitrate requirements. 


In the last decade, there had been numerous efforts to exploit the benefits of TVWS from both industries and academia. Microsoft and Google explored ways to leverage the TV band properties for wireless broadband Internet access in underserved communities~\cite{wifi_like_connectivity}. Microsoft has partnered with the governments of Jamaica, Namibia, Philippines, Tanzania, Taiwan, Columbia, UK, and the USA to provide wireless broadband Internet access over the TVWS~\cite{msrpartners}.
Similarly, various standard bodies are modifying existing standards and also  developing hardware platforms to accommodate this paradigm shift of opportunistically using TVWS and consulting with a
cloud service before using the medium. For example, the recently standardized IEEE 802.11af~\cite{IEE802_af} details techniques to enable Wi-Fi transmissions in the TVWS. IEEE 802.22~\cite{IEEE802_22} working  group intends to provide wireless broadband access  over TVWS. IEEE 802.19~\cite{IEE802_19} is aimed at enabling effective use of TVWS by the family of IEEE 802 standards. In parallel, the research community has been investigating techniques to make spectrum sensing low-cost~\cite{wiser1, wiser2}, more accurate~\cite{ws_dyspan08_kim, sensing2}, and to develop networking protocols~\cite{snow_ton, knows, winet}.

Along with providing wireless broadband Internet access, TVWS networking can subdue the shortcomings of existing technologies such as those based on IEEE 802.11~\cite{IEE802_11} and IEEE 802.15.4~\cite{ieee154m}, benefiting numerous other applications. Examples include sensing and monitoring applications (e.g, urban sensing~\cite{urban_sense1}, wildlife habitat monitoring~\cite{habitat1}, volcano monitoring~\cite{volcano1}, oil field monitoring~\cite{oilfield2}, civil infrastructure monitoring~\cite{civil1}), agricultural Internet-of-Things (IoT) applications (e.g., Microsoft Farmbeats project~\cite{farmbeats}, Climate Crop~\cite{climatecrop}, Monsanto~\cite{monsanto}, AT\&T project~\cite{atnt}), smart and connected community applications (e.g., public safety~\cite{elshafie2015survey}), real-time applications (e.g., industrial process control~\cite{processcontrol1}, civil structure control~\cite{civil2}, data center power management~\cite{capnet}), smart utility applications (e.g., advanced metering infrastructure~\cite{smartgrid1}), location-based services~\cite{mohapatra2005survey}, and transportation and logistics applications (e.g., connected-vehicles~\cite{lu2014connected}).

Networking over the TVWS faces a number of challenges for current and future adaptations. {\bf First}, interference between primary user (PU), i.e., TV station or other licensed user, and secondary user (SU), i.e., unlicensed devices, plays a significant role in the design of TVWS networking protocols. {\bf Second}, these protocols need to handle coexistence issues to facilitate homogeneous and heterogeneous applications. {\bf Third}, TVWS are mostly fragmented (i.e., discontinuous in frequency spectrum) and pose significant challenges in seamless operation of these protocols. {\bf Fourth}, their dynamic nature due to both temporal and spatial variations make it challenging to adopt them in any protocol, especially under mobility. {\bf Fifth}, They can be scarce in some places (mostly crowded urban areas) and these protocols need to be adaptive to other frequency bands when needed. 
{\bf Sixth}, proper security measures should be taken to ensure the communication safety in these protocols. 
\revise{In this paper, we perform a retrospective review of the protocols that have been built over the last decade and also the new challenges and the directions for future work. 
Additionally, we explore the TVWS regulations, characteristics, and the opportunities of the TVWS protocols for new application domains (e.g., IoT). While few recent surveys~\cite{elshafie2015survey,zhang2018tv, brown2014survey, wang2011emerging, fadda2015feasibility, han2015survey, 5345318, jingmin2017survey} provide insights on opportunities and challenges of networking in the TVWS, none of them provide a comprehensive review and comparison between those protocols.
To the best of our knowledge, this is the first comprehensive survey that reviews and compares existing networking protocols over the TVWS.}

The rest of the paper is organized as follows. Section~\ref{sec:tv_ws} provides TVWS regulatation. Section~\ref{sec:efforts} overviews standardization and industrial efforts for wireless broadband Internet access over the TVWS.
Section~\ref{sec:char} explores the characteristics of the TVWS. Section~\ref{sec:technologies} describes and compares different TVWS protocols. Section~\ref{sec:opportunities} describes the opportunities and Section~\ref{sec:challenges} identifies the key research challenges and directions for TVWS networking. Finally, Section~\ref{sec:conclusion} concludes our paper.

\section{Overview of the TV White Spaces Regulations}\label{sec:tv_ws}
In the US, TVWS lie within the VHF and lower UHF (54 -- 698 MHz) bands that correspond to TV channel indices 2 -- 51 (each with 6MHz channel spacing). Here, we discuss different regulations and standardization efforts on TVWS.

\subsection{Operational Regulations}

\paragraph{Licensed Operation}
Licensed access to TVWS is aimed mainly to enable the wireless broadband and mobile data services~\cite{FCC_first_order}. Specially, Qualcomm Inc., Fiber Tower Corporation, the Rural Telecommunication Group (RTG) Inc., Sprint Nextel Corporation, and T-Mobile USA encouraged FFC to allow licensed TVWS operation to assure no interference toward PUs from SUs. Also, licensing shall guarantee that all the innovations benefits may only be for the licensee, thus encouraging the big companies to invest on TVWS.

\paragraph{Unlicensed Operation}
On the other hand, unlicensed access to the TVWS may open up a whole new level of opportunities for low-power SUs. Existing unlicensed bands (e.g., 2.4/5 GHz) are already hosting many innovative products (e.g., Wi-Fi, ZigBee, BLE) due to no barriers in accessing unlicensed bands. 
Thus, unlicensed use of TVWS may flourish the innovations and definitely increase the chances of providing wireless broadband access to rural and Native American tribal areas due to its lower costs compared to the infrastructure investments in licensed operation.

\paragraph{Hybrid Operation}
The FCC, encouraged by Wireless Internet Service Providers Association, also allows accessing TVWS with hybrid or light licensing models, similar to the licensing rules in 3650 -- 3700 MHz band. As a viable alternative to exclusively licensed and/or unlicensed operations, hybrid licensing allows fixed SUs (e.g., Cellular towers) to register in a database and other SUs (personal/portable) not to use TVWS exclusively. 

\subsection{Unlicensed Access Regulations}
In this section, we discuss the unlicensed access regulations of the TVWS. Below, we first provide the unlicensed device classification by the FCC.

\subsubsection{Unlicensed SU Classification}

The FCC classified the SUs into two groups -- fixed and personal/portable.
\paragraph{Fixed Devices}
Fixed SUs can transmit with higher transmission (Tx) power and are generally placed in fixed outdoor locations. Fixed SUs are allowed to communicate with other fixed and personal/portable SUs. After acquiring a TVWS channel, a fixed SU has to send automatic periodic messages identifying itself to avoid interference with other PUs/SUs.  Functionally, they may provide commercial and/or non-commercial services. With higher Tx power, fixed SU may cover a geographical area of several kilometers. In the US, a fixed SU are allowed to use TV channels 2 -- 13 (54 -- 216 MHz) and 14 -- 51 (470 -- 698 MHz), except channels 3, 4, and 37.

\paragraph{Personal/portable Devices}
Personal/portable SUs operate with lower Tx power, may change location, and can be installed as Wi-Fi like cards in cellphone, laptop, computer, and wireless in-home devices. Personal/portable SUs have two operational modes: (i) controlled by a fixed SU or a personal/portable SU that has already determined its usable TVWS (Mode I), (2) independent, where a personal/portable SU determines its TVWS by itself (Mode II). Personal/portable SUs are allowed to communicate with both fixed and other personal/portable SUs. Personal/portable SUs may operate in TV channels 14 -- 51 (470 -- 698 MHz), except channels 20 and 37.

\subsubsection{Determining White Spaces}
According to the FCC regulations, there are three ways to determine TVWS in a location: geo-location database approach, control signal approach, and spectrum sensing. 

\paragraph{Geo-location Database Approach}\label{sec:database_approach}
In this method, a SU will first learn its geo-location via a professional device installer or GPS and then contact a licensed geo-location database hosted on the cloud (Internet) to determine the existing TVWS at its location. However, an analysis has to be done for each TVWS channel to determine whether the SU's location is in the interfering zone of any PUs or not~\cite{FCC_first_order}.

\paragraph{Control Signal Approach}
This method incorporates TVWS information in a control signal generated by an external source such as a fixed TV or radio broadcast station, a Commercial Mobile Radio Service (CMRS) base station, a Private Land Mobile Service (PLMRS) base station, or another unlicensed transmitter. A SU will listen to the control signals transmitted by the external sources. A SU is only allowed to transmit in a TVWS channel after it gets a positive identification of available TVWS channel from the control signals generated by a single or multiple sources.

\paragraph{Spectrum Listening/Sensing Approach}
A SU with the capability of sensing or listening to the TV signals can sense and determine the available TVWS channels in its geo-location. 
However, 
according to the FCC, a TV channel can be considered free if the energy level (e.g., signal power) in that TV channel is below  certain threshold values (see Table~\ref{tab:ws_tech_req}). These threshold values may also vary based on different physical attributes (i.e., device type and TV channel index).

The FCC in its first order in 2008~\cite{FCC_first_order} mandated that a fixed SU has to adopt both geo-location database
and spectrum sensing approaches to detect the TVWS at its location. On the other hand, a personal/portable SU may employ one of the two approaches: (i) it has to be under the control of another fixed or personal/portable SU with the capability of determining its location and accessing the geo-location database; (ii) it must be capable of determining its own location, accessing the geo-location database, and spectrum sensing.
However, in the FCC second order in 2010~\cite{fcc_second_order}, the mandatory rule of spectrum sensing for SUs has been lifted and kept as optional choice. Table~\ref{tab:ws_tech_req} summarizes all the regulatory parameters for networking in the TVWS.

\begin{table*}[htb!]
\centering
\scriptsize
\singlespace
\begin{tabular}{|c|c|c|c|}
\hline
\multirow{2}{*}{\textbf{\begin{tabular}[c]{@{}c@{}}Technical \\ Attributes\end{tabular}}} & \multirow{2}{*}{\textbf{Fixed Device}} & \multicolumn{2}{c|}{\textbf{Personal/portable Device}}                                                                                                                    \\ \cline{3-4} 
                                                                                          &                                        & \textbf{Mode I (Client mode)}                                                       & \textbf{Mode II (Independent mode)}                                                 \\ \hline
\textbf{Database Access}                                                                  & Necessary                              & Not necessary                                                                       & Necessary                                                                           \\ \hline
\textbf{Spectrum  Sensing/Listening}                                                      & Not necessary                          & Not necessary                                                                       & Not necessary                                                                       \\ \hline
\textbf{Allowed on TV Channels}                                                           & 2-35 except 3 \& 4                     & 21-35                                                                               & 21-35                                                                               \\ \hline
\textbf{Sensing Threshold}                                                                & -84 dBm (per 6 MHz)                    & -114 dBm (per 6 MHz)                                                                & -114 dBm (per 6 MHz)                                                                \\ \hline
\textbf{Transmission Power}                                                               & 30 dBm                                 & \begin{tabular}[c]{@{}c@{}}20 dBm, or 16 dBm \\ (Near protected Areas)\end{tabular} & \begin{tabular}[c]{@{}c@{}}20 dBm, or 16 dBm \\ (Near protected Areas)\end{tabular} \\ \hline
\textbf{Device Antenna Height}                                                            & \textless 30 meters (outdoor)          & N/A                                                                                 & N/A                                                                                 \\ \hline
\textbf{Antenna Gain}                                                                     & \textless 6 dBi                        & 0 dBi                                                                               & 0 dBi                                                                               \\ \hline
\textbf{Out-of-Band Emission Limits}                                                      & 4 W                                 & 100 mW                                                                           & 100 mW                                                                           \\ \hline
\end{tabular}
\caption{Technical Attributes of TV White Space Devices} \vspace{-0.3in}
\label{tab:ws_tech_req}
\end{table*}

\section{TV White Spaces Standardization and Industrial Efforts}\label{sec:efforts}
There have been multiple efforts from the  standardization bodies and industry to exploit the potential of the TVWS. We summarize those efforts as follows.
\paragraph{Standardization}
The usage of TVWS is standardized for wireless broadband Internet access (i.e., Wi-Fi-like connectivity) by the IEEE 802.11af wireless local area networks group~\cite{IEE802_af}. It focuses on developing cognitive medium access control (MAC) and physical layer (PHY) specifications. Similarly, IEEE 802.22 group focuses on developing cognitive MAC and PHY specifications for wireless regional area networks over the TVWS~\cite{IEEE802_22}. However, its main focus is to deal with the coexistence issues between PUs and SUs.
The usage of TVWS is also being standardized for low-power and low-rate networks. As such, IEEE 802.15.4m has adopted TVWS and provide MAC and PHY specifications for low-rate wireless personal area networks~\cite{ieee154m}, which shows great promise for limited-resource IoT applications. Also, IEEE 1900.4a proposed its amendments to enable mobile wireless services in the TVWS~\cite{ieee1900_4a}. It aims to provide architectural building blocks with optimized radio resources to enable distributed heterogeneous wireless access networks. Alongside, IEEE 1900.4.1~\cite{ieee1900_4_1} task group provides a detailed description of the building blocks for IEEE 1900.4a. 
IEEE 802.19.1 focuses on developing methods for coexistence of different unlicensed SUs/networks~\cite{IEE802_19}. Specifically, this task group specifies methods for coexistence in TVWS that are independent of any radio technologies. IEEE 1900.7 aims to develop a radio interface that incorporates MAC and PHY layers for dynamic access of TVWS for both fixed and personal/portable SUs~\cite{ieee1900_7}. \revise{Apart from the IEEE, ECMA-392~\cite{ecma}, proposed by Cognitive Networking Alliance (CogNeA) which embodies industry giants including Philips, Samsung, and Motorola~\cite{zhang2018tv}, focuses on developing MAC, PHY, and database technologies for SUs to avoid interference with PUs.}

\paragraph{Industrial Efforts Towards Wireless Broadband Internet Access}
Industry leaders, such as Microsoft and Google, explored ways to leverage the TVWS for wireless broadband Internet access in underserved communities in different parts of Africa~\cite{ms4afrika, broadbandmsr, broadbandgoogle}. The broadband access initiative started in 2013 as a project called {\em Microsoft 4Afrika}~\cite{ms4afrika}. Since then, from around the globe, Microsoft has partnered with the governments of Jamaica, Namibia, Philippines, Tanzania, Taiwan, Columbia, UK, and the USA to provide broadband Internet access over the TVWS~\cite{msrpartners}. Alone in the US, Microsoft, in collaboration with RADWIN -- a world leader in delivering high-performance broadband wireless access solutions~\cite{radwin}, has started the Microsoft {\em Airband Initiative} that includes 23 projects in rural areas to provide Internet connectivity to 20 million people of 15 states~\cite{msrairband, airbandnews}. 
The Airband Initiative is also working closely with several Internet service providers and other telecommunications companies, introducing innovative solutions for smart and connected communities. It is envisioned that during the second half of 2019, Airband Initiative will be introducing the innovative TVWS solutions to its partners.
Additionally, Microsoft is also trying to open up TVWS in India for broadband access~\cite{ws_india}. Companies like Carlson Wireless Technologies is also providing broadband Internet service to its users leveraging the TVWS~\cite{carlson}.

\section{Characteristics of the TVWS}\label{sec:char}
In this section, we discuss several characteristics of the TVWS.

\subsection{Spectrum Availability}
TVWS may provide a large number of channels for SUs in most places compared to the ISM bands.
With few restrictions (i.e., channel 37 is only for medical use, two channels are designated for wireless microphones), TV spectrum offers 47 channels, totaling $\approx 280$ MHz (6 MHz each). Depending on geo-location, a subset of these 47 channels may become TVWS. Statistically, rural and suburban areas offer a larger set of TVWS due to very few official TV stations~\cite{FCC_first_order}. Urban areas offer sufficient amount of TVWS, but lesser in number compared rural or suburban areas. 

\subsection{Wider Bandwidth}
TVWS provide diverse bandwidth choices for wireless communication. 
A device can operate on multiple, say $N$, consecutive channels, each 6 MHz wide, thereby using a bandwidth of $N*6$ MHz. 
Such diverse bandwidth selection may benefit many existing wireless technologies. 
In traditional WSNs such as those based on IEEE 802.15.4~\cite{ieee154m}, the maximum allowable bitrate is 250 kbps making them inefficient for many applications that require high data rate. 
For example, volcano monitoring~\cite{volcano1} applications require data being sent over wireless medium from seismometers and microphones at a very high sampling rate of 100 kHz. Also, they require a resolution of 24 bits per sample during transmissions. 
While IEEE 802.15.4 based networks thus suffer from severe limitations in meeting the QoS (Quality of Service) requirements of these applications, the TVWS show great promises for them.
A single TV channel may provide a bandwidth up to 21 Mbps~\cite{wifi_like_connectivity}, which can be increased by operating on several TVWS channels. 

\subsection{Propagational Characteristics}\label{sec:prop_char}

\paragraph{Longer Communication Range}
Wireless signals transmitted over the TVWS can reach longer distances due to its lower frequencies (54 -- 698 MHz). While communication range critically depends on the transmission power, lower frequency adds an extra beneficial factor to that. As per Friis' Transmission Equation~\cite{friis1946note}:
$P_{rx} = P_{tx}G_{tx}G_{rx}\big(\frac{c}{4\pi D_r f}\big)^2$,
where $P_{rx}$ and $G_{rx}$ are the receiver power and gain, respectively. $P_{tx}$ and $G_{tx}$ are the transmitter power and gain, respectively. $c$ is the speed of light. $D_r$ is the distance between transmitter and receiver and $f$ is the frequency that is used for communication. Looking closely, we can see that \emph{distance} is inversely proportional to the \emph{frequency}. 
Thus, confirming the relationship between lower frequencies and longer distances in wireless communication. 
In practice, a single hop communication over the TVWS ranges up to 8 km with a Tx power of 20 dBm~\cite{snow2}.


\paragraph{Obstacle Penetration}
TVWS spectra have improved obstacle penetration capability due to their longer wavelengths that range between 43 and 59 cm. Signals transmitted over the TVWS easily penetrate walls, allowing better non-line-of-sight connectivity unlike IEEE 802.15.4/IEEE 802.11 transmissions. White spaces hence are suitable for many environments that are challenging for current WSN technologies, e.g., those requiring wall penetration for radio through office buildings. Obstacle penetration is a severe problem in industrial environment where moving objects often block wireless communication. WirelessHART networks~\cite{wirelessHART} handle this through a high degree of redundancy where a packet is transmitted multiple times and through multiple paths hindering scalability. Transmissions over the white spaces can cover the entire network with negligible signal decay through walls and obstacles.  The lower frequency signals are less susceptible to human presence, multi-path and fading, and exhibit exceptional indoor penetration, and vary less over time. Hence, they hold potential for better indoor applications and localization~\cite{localization1}.

\begin{wrapfigure}{r}{0.5\textwidth}
\centering \vspace{-0.8in}
\includegraphics[width=0.47\textwidth]{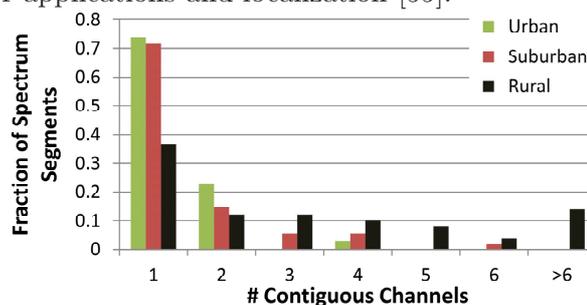} \vspace{-0.1in}
\caption{Spectrum fragmentation after US DTV transition in 2009~\cite{wifi_like_connectivity}.} \vspace{-0.25in}
\label{fig:frag}
\end{wrapfigure}

\subsection{Spectrum Fragmentation}
With the advent of analog to digital TV (DTV) broadcasting system, a TV channel bandwidth is narrowed down to a fixed 6 MHz width in the US~\cite{FCC_first_order} and other countries. 
While digital TV channels are relocated, depending on the geo-location (i.e., rural, suburban, and urban) and licensed/unlicensed usage, TVWS have naturally become fragmented, i.e., discontinuous in frequency spectrum.
As shown in Figure~\ref{fig:frag}, rural and suburban areas exhibit a much lower degree of fragmentation compared to urban regions~\cite{wifi_like_connectivity}.
Such fragmentation may disrupt seamless operations of the protocols that adopt TVWS. Moreover, it worsens when
protocols need to support mobility in the TVWS.

\section{TV White Space Networking Protocols}\label{sec:technologies}
In this section, we review the  existing TVWS protocols and provide a comparative study between them. We divide the protocols into three major categories: TVWS detection protocols, network architecture designs, and wireless broadband access protocols. The review of these protocols is followed by a table (Table~\ref{tab:comparison}) that summarizes their key features and differences.
\afterpage{
\begin{landscape}
\vspace*{\fill}
\begin{table}[htb!]
\centering
\scriptsize
\singlespace
\begin{tabular}{|c|c|c|c|c|c|c|c|c|c|c|c|c|}
\hline
\multicolumn{1}{|l|}{\multirow{2}{*}{}}                                       & \multicolumn{6}{c|}{\textbf{TVWS Detection Protocols}}                                                                                                                                                                                                                                                                                                                                                                                                                                                                    & \multicolumn{4}{c|}{\textbf{TVWS Network Architectures}}                                                                                                                                                                                                                                                                                                                                                 & \multicolumn{2}{c|}{\textbf{\begin{tabular}[c]{@{}c@{}}Wireless Broadband \\ Access Protocols\end{tabular}}}                                                                                                 \\ \cline{2-13} 
\multicolumn{1}{|l|}{}                                                        & \textbf{\begin{tabular}[c]{@{}c@{}}Waldo\\ \cite{sensing1}\end{tabular}}         & \textbf{\begin{tabular}[c]{@{}c@{}}FIWEX\\ \cite{sensing4}\end{tabular}}           & \textbf{\begin{tabular}[c]{@{}c@{}}SenseLess\\ \cite{geo1}\end{tabular}}   & \textbf{\begin{tabular}[c]{@{}c@{}}V-Scope\\ \cite{geo3}\end{tabular}}                                        & \textbf{\begin{tabular}[c]{@{}c@{}}HySIM\\ \cite{geo4}\end{tabular}}                     & \textbf{\begin{tabular}[c]{@{}c@{}}WISER\\ \cite{wiser1}\end{tabular}}                                        & \textbf{\begin{tabular}[c]{@{}c@{}}KNOWS\\ \cite{knows}\end{tabular}}                                         & \textbf{\begin{tabular}[c]{@{}c@{}}WINET\\ \cite{winet}\end{tabular}}                                          & \textbf{\begin{tabular}[c]{@{}c@{}}SNOW\\ \cite{snow_ton}\end{tabular}}                  & \textbf{\begin{tabular}[c]{@{}c@{}}CR-S\\ \cite{ahuja2008cognitive}\end{tabular}}                                            & \textbf{\begin{tabular}[c]{@{}c@{}}WhiteFi\\ \cite{wifi_like_connectivity}\end{tabular}}                                      & \textbf{\begin{tabular}[c]{@{}c@{}}WhiteNet\\ \cite{whitenet}\end{tabular}}                                      \\ \hline
\textbf{\begin{tabular}[c]{@{}c@{}}TVWS \\ Access\end{tabular}}               & \begin{tabular}[c]{@{}c@{}}Sensing\\ \& Local\\ Database\end{tabular} & \begin{tabular}[c]{@{}c@{}}Sensing \\ \& Local\\  Database\end{tabular} & \begin{tabular}[c]{@{}c@{}}Geo-\\ locaion\\ Database\end{tabular}   & \begin{tabular}[c]{@{}c@{}}Sensing,\\  Local\\  Database,\\ \& Geo-\\ location\\ Database\end{tabular} & \begin{tabular}[c]{@{}c@{}}Sensing\\  \& Geo-\\ location \\ Database\end{tabular} & \begin{tabular}[c]{@{}c@{}}Sensing \\ Local\\ Database,\\ \& Geo-\\ location\\ Database\end{tabular} & \begin{tabular}[c]{@{}c@{}}Sensing,\\ Local \\ Database,\\ \& Geo-\\ location\\ Database\end{tabular} & \begin{tabular}[c]{@{}c@{}}Sensing, \\ Local \\ Database,\\ \& Geo-\\ location\\ Database\end{tabular} & \begin{tabular}[c]{@{}c@{}}Geo-\\ location\\ Database\end{tabular}            & \begin{tabular}[c]{@{}c@{}}Sensing,\\  Local\\  Database,\\ \& Geo-\\ location\\  Database\end{tabular} & \begin{tabular}[c]{@{}c@{}}Sensing,\\ Local\\ Database,\\ \& Geo-\\ location\\ Database\end{tabular} & \begin{tabular}[c]{@{}c@{}}Sensing,\\ Local \\ Database,\\ \& Geo-\\ location\\ Database\end{tabular} \\ \hline
\textbf{PHY}                                                                  & N/A                                                                   & N/A                                                                     & N/A                                                                 & N/A                                                                                                    & N/A                                                                               & N/A                                                                                                  & OFDM                                                                                                  & N/A                                                                                                    & D-OFDM                                                                        & \begin{tabular}[c]{@{}c@{}}IEEE\\ 802.11a\end{tabular}                                                  & OFDM                                                                                                 & OFDMA                                                                                                 \\ \hline
\textbf{MAC}                                                                  & N/A                                                                   & N/A                                                                     & N/A                                                                 & N/A                                                                                                    & N/A                                                                               & N/A                                                                                                  & CSMA/CA                                                                                               & CSMA/CA                                                                                                & CSMA/CA                                                                       & CSMA/CA                                                                                                 & CSMA/CA                                                                                              & CSMA/CA                                                                                               \\ \hline
\textbf{\begin{tabular}[c]{@{}c@{}}Concurrent\\ Tx/Rx\end{tabular}}           & No                                                                    & No                                                                      & Yes                                                                 & No                                                                                                     & N/A                                                                               & No                                                                                                   & No                                                                                                    & Yes                                                                                                    & Yes                                                                           & No                                                                                                      & No                                                                                                   & Yes                                                                                                   \\ \hline
\textbf{\begin{tabular}[c]{@{}c@{}}Variable\\  User\\ Bandwidth\end{tabular}} & N/A                                                                   & N/A                                                                     & N/A                                                                 & N/A                                                                                                    & N/A                                                                               & N/A                                                                                                  & Yes                                                                                                   & No                                                                                                     & No                                                                            & No                                                                                                      & Yes                                                                                                  & No                                                                                                    \\ \hline
\textbf{\begin{tabular}[c]{@{}c@{}}Support\\ Fragmented\\ TVWS\end{tabular}}  & Yes                                                                   & Yes                                                                     & Yes                                                                 & Yes                                                                                                    & Yes                                                                               & Yes                                                                                                  & Yes                                                                                                   & Yes                                                                                                    & Yes                                                                           & Yes                                                                                                     & Yes                                                                                                  & Yes                                                                                                   \\ \hline
\textbf{\begin{tabular}[c]{@{}c@{}}Indoor/\\ Outdoor\end{tabular}}            & Outdoor                                                               & Indoor                                                                  & Outdoor                                                             & Outdoor                                                                                                & \begin{tabular}[c]{@{}c@{}}Indoor/\\ Outdoor\end{tabular}                         & Indoor                                                                                               & \begin{tabular}[c]{@{}c@{}}Indoor/\\ Outdoor\end{tabular}                                             & Indoor                                                                                                 & \begin{tabular}[c]{@{}c@{}}Indoor/\\ Outdoor\end{tabular}                     & \begin{tabular}[c]{@{}c@{}}Indoor/\\ Outdoor\end{tabular}                                               & \begin{tabular}[c]{@{}c@{}}Indoor/\\ Outdoor\end{tabular}                                            & \begin{tabular}[c]{@{}c@{}}Indoor/\\ Outdoor\end{tabular}                                             \\ \hline
\textbf{Mobility}                                                             & Yes                                                                   & No                                                                      & Yes                                                                 & Yes                                                                                                    & N/A                                                                               & Yes                                                                                                  & Yes                                                                                                   & Yes                                                                                                    & No                                                                            & Yes                                                                                                     & Yes                                                                                                  & Yes                                                                                                   \\ \hline
\textbf{Security}                                                             & Yes                                                                   & No                                                                      & No                                                                  & No                                                                                                     & No                                                                                & No                                                                                                   & No                                                                                                    & No                                                                                                     & No                                                                            & Yes                                                                                                     & No                                                                                                   & No                                                                                                    \\ \hline
\textbf{\begin{tabular}[c]{@{}c@{}}Application\\ Scope\end{tabular}}          & \begin{tabular}[c]{@{}c@{}}Outdoor\\ TVWS\\ Detection\end{tabular}    & \begin{tabular}[c]{@{}c@{}}Indoor \\ TVWS\\ Detection\end{tabular}      & \begin{tabular}[c]{@{}c@{}}Outdoor \\ TVWS\\ Detection\end{tabular} & \begin{tabular}[c]{@{}c@{}}Outdoor\\ TVWS\\ Detection\end{tabular}                                     & \begin{tabular}[c]{@{}c@{}}TVWS\\ Market\end{tabular}                             & \begin{tabular}[c]{@{}c@{}}Indoor\\ TVWS\\ Detection\end{tabular}                                    & \begin{tabular}[c]{@{}c@{}}Indoor/\\ Outdoor\\ TVWS\\ Networking\end{tabular}                         & \begin{tabular}[c]{@{}c@{}}Indoor\\ TVWS\\ Networking\end{tabular}                                     & \begin{tabular}[c]{@{}c@{}}WSN, IoT,\\  Smart\\ Utility, \\ etc.\end{tabular} & \begin{tabular}[c]{@{}c@{}}IoT,\\ Smart\\ Utility,\\ etc.\end{tabular}                                  & \begin{tabular}[c]{@{}c@{}}Wireless\\ Broadband\\ Access\end{tabular}                                & \begin{tabular}[c]{@{}c@{}}Wireless\\ Broadband\\ Access\end{tabular}                                 \\ \hline
\end{tabular}\vspace{5px}
\caption{Comparison between several TVWS protocols}
\label{tab:comparison}
\end{table}
\vspace*{\fill}
\end{landscape}
}

\subsection{TVWS Detection Protocols}
\subsubsection{Waldo (White Space Adaptive Local Detector)}
Waldo~\cite{sensing1} enables the low-cost devices to detect and opportunistically use the TVWS. 
It demonstrates that low-cost spectrum monitoring devices have good enough sensing capabilities. 
Its detection technique leverages locally measured signal features and location. 
In short, it exploits the crowdsourced local spectrum measurements data from multiple low-cost sensing devices.

\begin{wrapfigure}{r}{0.5\textwidth}
 \vspace{-0.3in}
\includegraphics[width=0.45\textwidth, height=5.67cm, right]{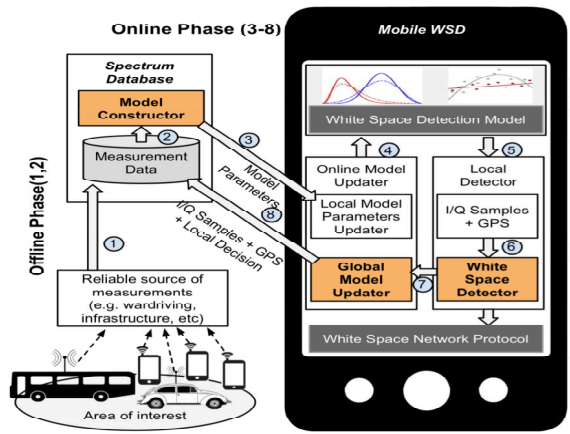} \vspace{-0.15in}
\caption{The flow of operation in Waldo~\cite{sensing1}} \vspace{-0.15in}
\label{fig:waldo}
\end{wrapfigure}

Waldo has two basic modules: a local centralized database and the TVWS devices. Also, it operates in two phases: offline and online. Its operational flow is shown in Figure~\ref{fig:waldo}.
In {\em offline phase}, the central database module collects spectrum characteristics and associated locations from different low-cost dedicated sensing devices. Spectrum information from a various number of devices is fused together to construct local models for TVWS availability at distinct locations.
In {\em online phase}, TVWS devices collect location-specific model information from the database and correlate with their locally measured spectrum characteristics. Based on the correlation, a device decides on the TVWS availability for opportunistic usage. The TVWS devices also feedback their local spectrum information to the database. 
As shown in Figure~\ref{fig:waldo}, the database module has a {\em model constructor} that uses a {\em binary classifier} to indicate whether a TV channel is a TVWS or not. 
Additionally, a {\em model updater} updates the location-based information.

\subsubsection{In-band Spectrum Sensing in Cognitive Radio Networks}

\begin{wrapfigure}{r}{0.5\textwidth}
\centering \vspace{-0.25in}
\includegraphics[width=0.45\textwidth]{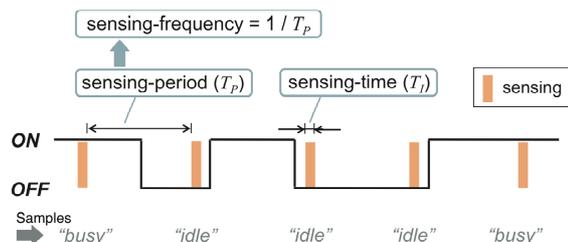}
\caption{In-band spectrum sensing (ON: occupied) in~\cite{sensing2}.} \vspace{-0.15in}
\label{fig:inband}
\end{wrapfigure}

To properly model the coexistence between PU and SU, the work in~\cite{sensing2} proposed an efficient and effective in-band spectrum sensing-scheduling technique. \revise{In-band channels refer to the TVWS channels that are currently occupied by the SUs. In-band spectrum sensing, adopted in the SUs, is essential to detect the returns of the PUs so that a SU can vacate the in-band channels to avoid interference to the PUs. The work in~\cite{sensing2} presents a periodic in-band spectrum sensing protocol (shown in Figure~\ref{fig:inband}) that optimizes both sensing interval and latency while minimizing sensing overhead. It also reduces the false positives and the false negatives in TVWS detection.
In-band sensing-scheduling algorithm incorporates {\em fast sensing} (calculates average received signal strength on a channel over 1 ms) or {\em fine sensing} (uses feature detection such as pilot energy detection in the DTV transmissions with a sampling frequency of 70 kHz) while minimizing the overall sensing overhead}. 
In fast sensing, the algorithm takes the minimum amount of time. 
However, fast sensing is highly vulnerable to channel noise or co-channel interference. In contrast, fine sensing needs relatively long time compared to fast sensing and provides a higher degree of sensing accuracy.
Depending on sensing frequency and QoS at different times, the in-band sensing scheduling algorithm will employ either fast sensing or fine sensing.


\subsubsection{A TV White Space Spectrum Sensing Prototype}
The work in~\cite{sensing3} presents a real-time TVWS spectrum sensing prototype that detects signals from ATSC digital TV devices, NTSC analog TV devices, and wireless microphones in a single integrated platform. It solves the fading channel problem in ATSC and NTSC using a spatial diversity technique. It deals with the weaker signals from wireless microphones using an advanced sensing system that accounts for the surrounding noise. 
The prototype incorporates two configurable receiver antennas for spatial diversity. Any single antenna or both can be used for sensing. The prototype has a single RF receiver chain. Thus, in {\em diversity mode},  the receiver interchanges between antennas with a time interval, called the \emph{quiet time}. A quite time is set up generally in two ways: long quite time or sequence of shorter quite times. This prototype especially uses a sequence of shorter quiet times, each approximately 7 ms. After receiving the baseband samples from sensing, the prototype applies a Fast Fourier Transform (FFT) algorithm.
The absolute squared output from the FFT algorithm is scaled, accumulated, and sent to ATSC, NTSC, and wireless microphone sensors. 
Based on processing from these specific sensors, the prototype decides whether the sensed signal is from ATSC, NTSC, or wireless microphone and filters out the TVWS.

\subsubsection{FIWEX (Cost-efficient Indoor White Space Exploration)}
FIWEX~\cite{sensing4} studies the temporal and spatial features of the indoor TVWS spectrum. It uses a limited number of RF-sensors deployed inside a building that enables it to accurately determine highly accurate indoor TVWS availability at low cost. It determines whether a TV channel is vacant or not by comparing the received signal strength (RSS) in that channel with a threshold value. Thus, if the RSS in a TV channel is higher than the threshold value, that TV channel is deemed locally occupied, otherwise not. In fact, it can operate with the devices that can support a threshold level of -114 dBm (as per FCC~\cite{FCC_first_order}).

\begin{wrapfigure}{r}{0.5\textwidth}
\centering \vspace{-0.5in}
\includegraphics[width=0.5\textwidth]{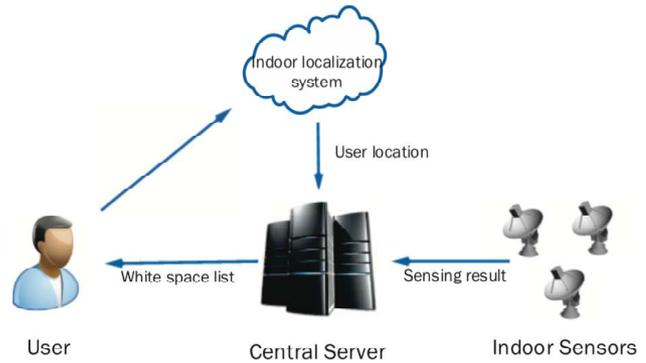} \vspace{-0.3in}
\caption{System architecture of FIWEX~\cite{sensing4}} \vspace{-0.15in}
\label{fig:fiwex}
\end{wrapfigure}
The FIWEX mechanism consists of long-time and short-time sensing. In {\em long-time sensing}, it listens to specific locations altogether (one receiver for each location) for a longer period of time. In contrast, it chooses few candidate locations in {\em short-time sensing} and moves a single receiver to gather RSS information. Based on long-time and short-time sensing data, it finds the strong TV channels that are mostly occupied and TVWS that are available at different times at different locations. The FIWEX system, as illustrated in Figure~\ref{fig:fiwex}, has a central server and a real-time sensing module. The {\em central server} utilizes the data from real-time sensing modules (e.g., RF-sensors) and maintains historical records of TV channels and their relative location-based TVWS competency information. The system architecture of FIWEX is depicted in Figure.


\subsubsection{SenseLess}
A database-driven TVWS protocol called SenseLess has been proposed to safely and efficiently operate a network in the TVWS~\cite{geo1}. SeneseLess takes into account the up-to-date database inputs about the TV channel incumbents, TV channel signal propagation modeling, and a content dissemination process to ensure interference-free and scalable white space networking.

\begin{figure}[!htbp]
    \centering \vspace{-0.3in}
      \subfigure[Architectural overview\label{fig:senseless1}]{
    \includegraphics[width=0.37\textwidth, height=3cm]{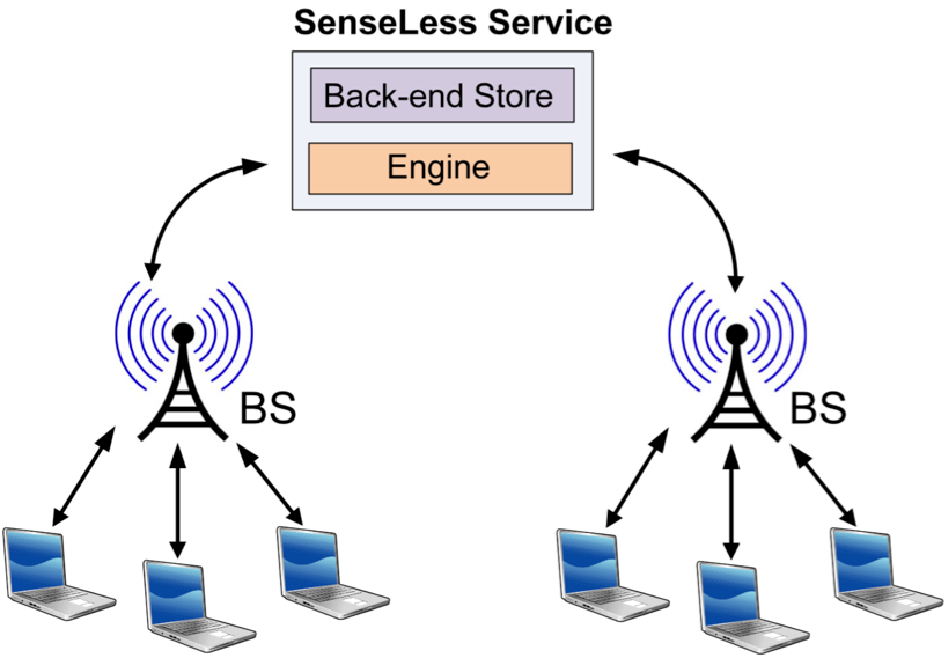}
      }\hfill
      \subfigure[Internal components\label{fig:senseless2}]{
        \includegraphics[width=.5\textwidth]{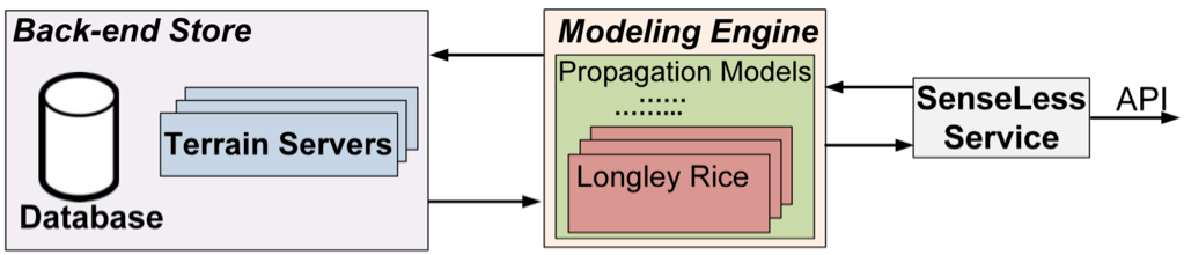}
      } \vspace{-0.15in}
    \caption{The SenseLess system architecture~\cite{geo1}.} \vspace{-0.15in}
    \label{fig:snow_arch}
 \end{figure}
The design of SenseLess is infrastructure based. It has several base stations (BSs) and clients, where each client is associated with a BS, as shown in Figure~\ref{fig:senseless1}. BSs are logically connected to a SenseLess service. For a given location, the {\em SenseLess service} yields the available TVWS. In SenseLess, neither the BSs nor the client employ any spectrum sensing protocol. It solely depends on the SenseLess service to learn about the TVWS at their locations. The service provides a subset of APIs and can operate in two different modes: sole location input or publish-subscribe. {\em Sole location input mode} provides with the TVWS information for a given location. In {\em publish-subscribe mode}, a BS or client node (via BS) can subscribe to the SenseLess service. The SenseLess service tracks the changes in TVWS at the subscribers' locations. If the TVWS change in a subscribed location, it will fire an event to the BS of a client node or a BS itself. Thus, BSs are always connected to the SenseLess service.
As shown in Figure~\ref{fig:senseless2}, it has two components: back-end store and SenseLess engine.
The {\em back-end store} has a database and a terrain-servers component. The database stores information (e.g., location, channel, height, transmit power) about all the TV stations, wireless microphones, and the subscribed BSs/client nodes. The terrain-servers provide terrain elevation data at any given point of the device's location. 
On the other hand, {\em SenseLess engine} collects all the parameters that are available in the database and determines the TVWS. 


\subsubsection{V-Scope (Vehicular Spectrum Scope)}
The work in~\cite{geo3} presents V-Scope, a vehicle-assisted spectrum measurement framework. It enhances the performance and utilization of the already existing database-based TVWS detection protocols by incorporating sensing-aided data into the database.
It proposes to attach RF-sensors on vehicles and collect and analyze data from those sensors to better estimate the TVWS availability. 
It comprises five workarounds that contribute to enhance the traditional TVWS database with measurements from sensors mounted on moving vehicles. {\bf First}, it detects the PUs and SUs based on measurements on different TV channels.
{\bf Second}, all the measurements are clustered into different groups after being sent to a central server. {\bf Third}, each TV channel propagation models are refined. {\bf Fourth}, SUs are localized if their locations are not already registered into the database. {\bf Fifth}, it models the power leakage to estimate the TVWS that are adjacent to the PUs.

\subsubsection{WISER (White Space Indoor Spectrum Enhancer)}\label{label:wiser}
The work in~\cite{wiser1} presents an indoor TVWS exploration system where SUs do not sense TVWS. Traditional conservative database-based approach misses out to identify a significant portion of indoor TVWS. WISER proposes cost-effective techniques to undiscovered indoor TVWS. It outperforms outdoor-sensing-only, one-time-profiling-only, and sensor-all-over-the-place methodologies.

\begin{wrapfigure}{r}{0.5\textwidth}
\centering \vspace{-0.3in}
\includegraphics[width=0.5\textwidth]{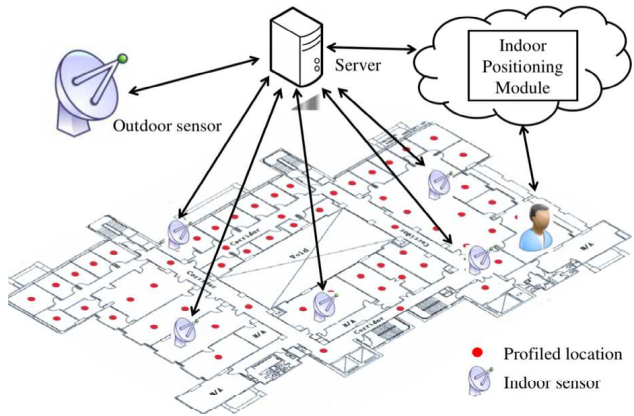} \vspace{-0.3in}
\caption{Architecture of WISER~\cite{wiser1}.} \vspace{-0.15in}
\label{fig:wiser}
\end{wrapfigure}
WISER consists of three modules: real-time sensing module, database, and indoor positioning module, as shown in Figure~\ref{fig:wiser}. 
The {\em real-time indoor sensing module} reports the availability of the TVWS to the database. 
It first conducts one-time spectrum profiling by sensing at sufficient indoor locations and builds a correlation between TV channels and the locations. The location that is not profiled is assumed to have the same correlation as the nearest profiled location. After profiling, TV channels are grouped into two categories: strong and weak-to-normal. The grouping also includes the permanent TVWS channels from TVWS database. The intuition is that the weak-to-normal TV channels have much higher chance to become TVWS. The profiling data also helps to build a channel-location clustering. 
Finally,
based on channel-location clustering, indoor sensors are positioned to contribute to WISER database. Hence, the real-time sensing module is mostly responsible for its correctness. A SU that wants to access TVWS first determines its location using the {\em indoor positioning module} and then queries the database.

\subsubsection{HySIM (Hybrid Spectrum and Information Market)}
The work in~\cite{geo4} presents a game theory-based protocol for a TVWS network that incorporates interactions between database operator, spectrum licensee, and the SUs. HySIM has a three-layered hierarchical architecture. At higher layer, the database and licensee negotiate the commission fee that licensee pays for spectrum market usage. In particular, the database uses the {\em revenue sharing scheme}~\cite{harsanyi1977rational} that fixes percentages of revenue sharing between them. In middle layer, the database and the licensee compete to sell information about the TVWS or other TV channels to the SUs. 
In particular, the licensee and the database decide on the price of the licensed and unlicensed TV channels, respectively, based on {\em supermodular game theory}~\cite{topkis2011supermodularity}.
In the lower layer, the SUs decide whether they should buy exclusive access to the channels from the licensee or the information about the TVWS from the database. based on the licensed and unlicensed TV channel price provided by the licensee and the database, respectively.

\subsection{Network Architecture Designs}

\subsubsection{KNOWS \revise{(Kognitive Network over White Spaces)}}\label{subsec:knows}
KNOWS proposes a robust TVWS detection and networking scheme where the SUs do not interfere the operation of the PUs~\cite{knows}. Its MAC protocol allows the SUs to opportunistically access and share variable bandwidth TVWS spectrum. Such design effectively can increase the overall throughput.
Its PHY layer consists of four modules: reconfigurable radio, scanner radio, GPS receiver, and x86 embedded processor. The {\em reconfigurable radio} comprises a commodity IEEE 802.11g~\cite{IEE802_11} card and a frequency synthesizer that can convert received signals to prespecified frequencies. The bandwidth choices of the reconfigurable radio are 5, 10, 20, and 40 MHz to operate in the TVWS with output powers between -8 dBm to 23 dBm. The reconfigurable radio has a 100 $\mu$s of time overhead for adjusting its parameters. The {\em scanner radio} periodically scans each TV channel for 10 ms in every 30 minutes to determine the TVWS. The scanner measures the received signal power in the TV channels at the resolution of 3 kHz. Most of the time, the scanner works as a receiver operating on the control channel located at 902 -- 928 MHz band. The {\em GPS receiver} determines the physical location and looks for available TVWS in case any white space database is accessible. Finally, the {\em x86 embedded processor} controls all the other three modules and interactions between them. 

The KNOWS MAC layer has two functions: collaborative sensing and spectrum reservation. In {\em collaborative sensing}, all one-hop neighboring SUs collectively learn about their TVWS by sending periodic beacon messages every 100-200 ms. 
To transfer data, sender and receiver nodes contact each other via the control channel and decide on the TVWS channel and desirable bandwidth to use by a three-way handshaking. Both sender and receiver contend for the control channel using carrier sense multiple access/collision avoidance (CSMA/CA) with a random back-off. Before starting the data transmission, the sender node reserves a TVWS channel by sending a spectrum reservation command packet, called {\em Data Transmission Reservation} (DTS). After sending the DTS, the sender and receiver initiate the exchange of data without any back-offs. A node in the network can build a log of the usage of the TVWS channels by listening to DTS commands from its neighbors.

\subsubsection{WINET \revise{(White Space Indoor Network)}}
WINET is the first indoor multi-AP TVWS network design~\cite{winet}. Considering the TVWS temporal variation, spatial variation, and fragmentation it optimizes the AP locations, AP association, and spectrum allocation, which lead to increase in the network coverage area, overall system throughput, and the fairness among its users.

The WINET architecture has three key components: WISER, APs, and demand locations. It adopts WISER (Section~\ref{label:wiser}) to identify more TVWS in indoor locations compared to the traditional database approach. Each AP and client has multiple cognitive radios. Any demand location of WINET can have multiple client devices, each capable of simultaneously contact with multiple APs though its multiple radios. Thus, multiple APs are deployed to cover all the demand locations. All the APs and client devices adopt the CSMA/CA MAC protocol. WINET addresses several key challenges to enable the multi-AP based indoor white space networking. {\bf First}, it provides a way to optimize the total number and placements of the APs that can maximize the system throughput and fairness between client devices. {\bf Second}, WINET compensates for the TVWS temporal variation, and spatial variation, and fragmentation. {\bf Third}, the association between APs can be significantly challenging since client devices may need to simultaneously contact with multiple APs. As such, WINET formulates two optimization problems: maximizing the average demand fairness problem (MDF) and spectrum allocation and AP association problem (SAP). MDF deals with the fairness between clients and SAP deals with AP's spectrum allocation and association. It provides a polynomial time solution for MDF and proves that SAP is NP-complete. 

\subsubsection{CR-S (A Cognitive Radio System)}
Using the IEEE 802.11a~\cite{IEE802_11} MAC, the work in~\cite{ahuja2008cognitive} presents a cognitive radio network architecture over the TVWS. The proposed network comprises a cognitive AP and various cognitive mobile stations.
For detecting TVWS, CR-S adopts both database and sensing based protocols. Such hybrid approach provides more protection against the incumbents and SUs.

The cognitive AP decides on the available TVWS to operate on and lets the cognitive mobile stations know via multiple control channels. Also, in-band signaling is adopted to enable network management and control. The cognitive AP is equipped with two transceivers: one is for dedicated sensing of the TV channels and the other is for dedicated communications between the AP and the cognitive mobile stations. Additionally, the AP consists of several modules: a policy engine, geo-location module, CE-UMAC interface, sensing engine, AP cognition controller, sensing information manager, and a graphical user interface (GUI). 
The AP communicates with the IEEE 802.11a MAC via a set of abstractions provided by the {\em CE-UMAC interface module}. The {\em AP cognition controller} is responsible for selecting the TVWS channel to operate on, keeping a rank of the TVWS channels, detecting and recovering from the in-band incumbents and SUs, and assessing the link qualities. On the other hand, each cognitive mobile station is equipped with a single transceiver. Also, a cognitive mobile station consists of a mobile station cognitive controller and a CE-UMAC interface. The function of CE-UMAC interface in mobile stations is the same as it is in the AP. 
The {\em mobile station cognitive controller} responds to all the control commands sent by the AP cognition controller.

\subsubsection{SNOW (Sensor Network Over White Spaces)}
SNOW is a highly scalable LPWAN technology operating in the TVWS~\cite{snow,snow2, snow3, snow_ton, snow_p2p, snowcots}. It supports asynchronous, reliable, bi-directional, and concurrent communication between a BS and numerous nodes. Due to its long-range, SNOW forms a star topology allowing the BS and the nodes to communicate directly. The BS is Internet-connected and line-powered and the nodes are power-constrained and do not have Internet access. To determine TVWS availability, the BS queries a geo-location database. A node depends on the BS to learn its TVWS availability. Each node is equipped with a half-duplex radio. To support simultaneous uplink and downlink communications, the BS uses a dual-radio architecture for reception (Rx) and transmission, respectively.

The SNOW PHY layer uses a distributed implementation of orthogonal frequency division multiplexing (OFDM) called D-OFDM. {\em D-OFDM} is designed for multi-user access. The BS operates on a wideband channel split into orthogonal narrowband subcarriers. Each node is assigned a single subcarrier when the number of nodes is no greater than the number of subcarriers. Otherwise, a subcarrier is shared by more than one node. 
SNOW supports two modulation techniques - ASK and BPSK, supporting different bitrates. Additionally, SNOW is capable of exploiting fragmented white space spectrum.
The nodes use a lightweight CSMA/CA MAC protocol similar to TinyOS~\cite{snow2}. Additionally, the nodes can autonomously transmit, remain in receive mode, or sleep. A node runs clear channel assessment (CCA) before transmitting. If its subcarrier is occupied, the node makes a random back-off in a fixed congestion back-off window. After this back-off expires, the node transmits immediately if its subcarrier is free. Then the node repeats this operation until it sends the packet and gets the acknowledgment (ACK).

\subsection{Wireless Broadband Access Protocols}

\subsubsection{Toward Enabling Broadband for a Billion Plus Population with TVWS}

The work in~\cite{broadband1} presents a broadband access-network topology using the TVWS in rural India. While India has the second largest telecommunication infrastructure, the majority of the rural areas in India lacks network (especially, Internet) connectivity. To provide wireless broadband Internet, multiple Wi-Fi clusters, covering several rural areas, connect to a fiber network over the TVWS. A broadband access-network may be possible in rural India by lengthening the Internet coverage from a rural PoP (Point of presence), provided by BharatNet~\cite{broadband1}. To extend the Internet reachability, TVWS are utilized to connect a PoP with an optical fiber point to Wi-Fi access points (APs). Using TVWS, non-line-of-sight and long-distance communications are possible. Rural people may connect to the Wi-Fi APs via 2.4 GHz band, where each Wi-Fi AP is equipped with a UHF band module. The UHF module backhauls data to/from the Wi-Fi mesh network. 

\subsubsection{WhiteFi (White Space Networking with Wi-Fi Like Connectivity)}\label{sec:whitefi}

The work in~\cite{wifi_like_connectivity} presents WhiteFi, built on KNOWS platform (Section~\ref{subsec:knows}), where Wi-Fi like connectivity is enabled in the TVWS. 
Such connectivity needs to address temporal variation, spatial variation, and fragmented spectrum of the TVWS. In fact, the design of
WhiteFi is validated based on extensive characterization and comparison (against 2.4 GHz Wi-Fi) of temporal variation, spatial variation, and spectrum fragmentation of TVWS.
It has three basic components: spectrum assignment algorithm (SAA), signal interpretation before Fourier Transform (SIFT) technique, and chirping protocol. 
{\em SAA} discovers the available TVWS for Wi-Fi operation by analyzing the spatial diversity and fragmentation of the TV spectrum. Additionally, 
SAA is highly adaptive to dynamics of TVWS and can assign variable channel-bandwidth to the clients (5, 10, 20 MHz). 
The {\em SIFT} performs the time-domain signal analysis of the TVWS spectrum and enables Wi-Fi AP and associated center frequency discovery. 
The {\em chirping protocol} deals with the client's disconnection from the Wi-Fi access point (due to unavailability of the TVWS) by reserving a separate 5 MHz backup channel. The information about the backup channel is available within each Wi-Fi beacon.

\subsubsection{WhiteNet}
\begin{wrapfigure}{r}{0.5\textwidth}
\centering \vspace{-0.6in}
\includegraphics[width=0.5\textwidth]{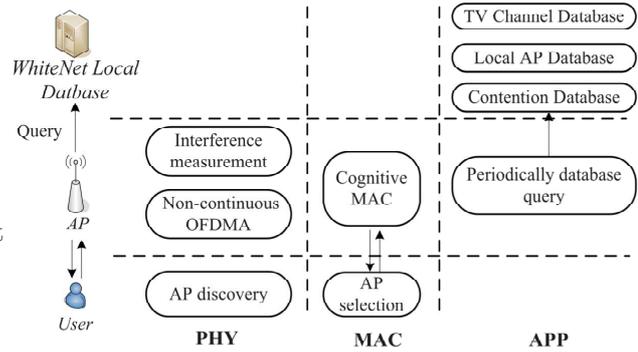} \vspace{-0.35in}
\caption{System overview of WhiteNet~\cite{whitenet}.} \vspace{-0.17in}
\label{fig:whitenet}
\end{wrapfigure}
The work in~\cite{whitenet} presents a database-assisted multi-AP TVWS broadband access-network architecture called WhiteNet. WhiteNet is similar to the WhiteFi (Section~\ref{sec:whitefi}) but differs mostly in the number of APs and provide a Wi-Fi like connectivity in a larger area. 
Due to the adoption of multiple APs, it is subject to inter-AP interference and spectrum allocation problem between the APs. It proposes an AP discovery method that helps a user to efficiently identify the center frequency and spectrum allocation of the corresponding AP.

As shown in Figure~\ref{fig:whitenet}, the WhiteNet architecture has three modules: local database, TVWS detection algorithm - called B-SAFE, and AP discovery. The {\em local database} conforms to the FCC database requirements. Apart from providing the available TVWS information, the local database also stores each AP location and their spectrum allocation information, interference of each TVWS channel on each AP, and inter-AP interference relationships. The local database has a submodule called {\em Contention Database} that resolves the TVWS channel contention between APs. The {\em B-SAFE} algorithm in each AP helps to determine TVWS in a distributed fashion by taking into account several constraints such as the interference of each TVWS channel on that AP, the corresponding inter-AP interference, fragmentation, and the area of coverage. The {\em AP discovery module} is directly associated with the users and helps to effectively and quickly identify the surrounding AP's center frequency and operational bandwidth. As part of the AP discovery module, 
each AP sends a periodic beacon message on the leftmost chunk of its acquired TVWS and encode its center frequency and bandwidth information. A user listens to these beacon messages to discover each AP. Additionally, a user performs a {\em Discovery and Cancellation} action, where it discovers one AP after another. After discovering all the APs, a user connects with the AP that can provide maximum overall utilization in terms of bitrate and  signal-to-interference-plus-noise ratio.

\subsubsection{Performance Analysis of a Wi-Fi Like Network Operating in TVWS}
 The work in~\cite{simic2012wi} presents a quantitative analysis of Wi-Fi-like networks those operate in the TVWS to provide wireless broadband access. Considering the inter-AP interference and dynamics of TVWS in urban and rural areas, it analyzes the achievable range and downlink throughput of Wi-Fi-like networks. 
 The performance analysis has been done in three scenarios: outdoor-urban, indoor-urban, and outdoor-rural areas, with a coverage area of (2x2) sq. km, (0.5x0.5) sq. km, and (5x5) sq. km, respectively. In each scenario, the locations of the APs are modeled using homogeneous Poisson Point Process~\cite{miles1970homogeneous}. The density of the process is set to 12.5 APs/sq. km, 0.25 APs/sq. km, and 125 APs/sq. km for outdoor-urban, indoor-urban, and outdoor-rural, respectively. The propagational models used for different scenarios are similar to~\cite{rappaport2002wireless}. To coexist, multiple Wi-Fi APs adopt the CSMA/CA MAC protocol. Based on all these models, the work in~\cite{simic2012wi} quantitatively showed that the range difference between traditional Wi-Fi AP operating on a 24 MHz wide TVWS channel in the outdoor-urban area stays between 67 m to 130 m. However, the downlink data rate can be reduced by 23\% even when using a 20\% wider bandwidth than traditional Wi-Fi settings. For indoor-urban scenario, the range limit can be extended up to 10 -- 20 m, while a similar downlink rate as traditional Wi-Fi network can be provided. The Outdoor-rural scenario shows similar characteristics as the indoor-urban scenario because of the lesser number of interference issues compared to the outdoor-urban scenario.

\section{Opportunities of TV White Space Networking}\label{sec:opportunities}

In this section, we provide a detailed application scope of TVWS networking. 

\paragraph{Sensing and Monitoring Applications}
In practice, wide-area sensing and monitoring applications are built on top of IEEE 802.15.4~\cite{ieee154m} and IEEE 802.11~\cite{IEE802_11}. 
However, due to their shorter communication range (e.g., 30-40 meters in a single hop), they create complex multi-hop mesh network topologies to cover larger geo-location area.
For examples, to cover the 1280 meters (the main span) of the Golden Gate Bridge in San Francisco, California, a WSN needs to be deployed with at least 46 hops~\cite{civil1} and it takes nearly 10 hours to collect data from all the sensors.
Intrinsically, protocols for such network will be very complicated, energy and time-consuming. 
Recently, such sensing and monitoring applications started adopting LPWAN technologies~\cite{dali_survey} that operate on the sub-1GHz band (902-928 MHz) and reduce them to single-hop networks. Similar to the sub-1GHz band, TVWS have same capabilities. Moreover, depending on the geo-location, TVWS can offer 5-10x more free spectrum. Thus, making it possible to host a great deal of wide-area sensing and monitoring applications including urban sensing~\cite{urban_sense1}, wildlife, environment, and habitat monitoring~\cite{wildlife1, environmental1, habitat1}, volcano monitoring~\cite{volcano1}, and oil field monitoring~\cite{oilfield2}.

\paragraph{Agricultural IoT Applications}
Agricultural IoT applications demand reliable communication between numerous sensors deployed in crop fields and farmers' warehouse~\cite{agriculture1}. 
The United Nations has envisioned to double the food production by 2050 to meet the future demand by the growing population~\cite{un_food}.
To efficiently and effectively farming, farmers should decide on the important and critical environmental factors and crop requirements on demand. As such, soil nutrients, effective fertilizers, water level in soil, seeds planted, temperature of stored food and materials should be monitored exclusively, all the time. To meet such demand, numerous sensors should be deployed and data should be collected to take effective data-driven decisions. 
The communication protocols between sensors can be built on TVWS to confirm longer-range communication, higher bandwidth, less interference, and  energy efficiency. The
Microsoft Farmbeats project~\cite{farmbeats} have already started adopting TVWS for building communication protocols for agricultural IoT applications. 

\paragraph{Applications for Smart and Connected Communities}
Protocols can be built on TVWS to enable applications that can enhance sustainability, quality of life,  health, safety, and economic prosperity of communities in both urban and rural areas~\cite{scc1}. 
The smart and connected community applications are envisioned to address the needs of past, present, and future of the community. Thus, ensuring the preservation and revitalization, livability, and sustainability of the community~\cite{scc2}. 
The enabling network technologies for smart and connected communities can be built over the TVWS. 
In fact, different attempts have already been made to enable wireless broadband access for rural areas ~\cite{broadbandmsr, broadbandgoogle}, enterprise, campus, municipal, and public safety~\cite{elshafie2015survey}, etc.

\paragraph{Real-Time Applications}
TVWS have the potential to support large-scale wireless cyber-physical systems (CPS) applications. 
Real-time CPS incorporates feedback control loops that depend on hard/soft real-time wireless communications between sensors to controllers and controllers to actuators. 
In contrast to traditional WSNs, i.e., 802.15.4~\cite{ieee154m},  CPS applications aim to achieve a best effort communication paradigm. 
Following the demand of CPS, industrial WSN protocols, i.e., WirelessHART~\cite{wirelessHART} have evolved over time. 
However, to meet the escalating demands of industrial Process control~\cite{processcontrol1}, Civil structure control~\cite{civil2}, thousands of sensor nodes will need to connect to the controller via gateways~\cite{wirelessHART}. In the WirelessHART~\cite{wirelessHART} standard, connecting thousands of sensors in an industrial setting will require several gateways with physical wires, thus losing the benefits of wireless. Also, due to the time synchronization requirements in the nodes and centralized network architecture, future real-time CPS applications will suffer from scalability issue. 
Thus, real-time CPS applications can be benefited greatly by adopting these protocols.   

\paragraph{Smart Utility Applications}
The ubiquitous smart utility network (SUN) for efficient management of electricity, natural gas, water, and sewage can be hugely benefited by adopting TVWS protocols. 
One of the key enabling features of SUN is the advanced metering infrastructure (AMI)~\cite{smartgrid1}. The benefits of AMI is twofold: (i) it provides monitoring, command, and control for the service providers; (ii) it helps in measurement, data collection, and analysis at consumers end. 
AMI tools at the provider and consumer ends are usually far apart from each other and need to exchange control messages and metering data. Such messaging requires efficient and inexpensive communication protocols that are highly scalable. 
Typically, SUN operates in unlicensed sub-1GHz (902-928 MHz) in the US. With several other LPWAN technologies (e.g., LoRa~\cite{lora}) being developed and targeted for IoT applications in the sub-1GHz band, it is only a matter of time that frequency band gets overly congested and prone to severe inter-technology interference~\cite{dali_survey}.
Thus, by adopting protocols built on TVWS will greatly benefit SUN applications. At the same, the communication range between the service provider and consumers can be significantly increased.

\vspace{-0.3in} 
\revise{\paragraph{Location-based Services}
TVWS networking may enable several location-based services including social event recommendations, assistive healthcare systems, alert systems, etc., especially in indoor locations such as malls, hospitals, airports, and subways. As of today, indoor localization is enable by several protocols such as ultra-wide band, ultrasonic, Wi-Fi, Zigbee, Bluetooth, and frequency modulation broadcast~\cite{zhang2018tv}. While the Wi-Fi signal fingerprint-based indoor localization techniques are mostly popular due to its ubiquitous availability, it suffers several limitations due to its high obstacle penetration loss. Additionally, reflection and diffraction  of Wi-Fi signals at both 2.4 and 5 GHz bands make it very difficult to scale indoor localization. As discussed in Section~\ref{sec:char}, TVWS spectrum have benefiting propagational characteristics and thus hold the potential to host scalable indoor localization techniques and enable several location-based services.}

\vspace{-0.3in} 
\revise{\paragraph{Transportation and Logistics}
For real-time tracking of inventories, products, and services from source to destination, business owners nowadays attach sensor tags to the transportation vehicles such as airplanes, trucks, and unmanned aerial vehicles (UAVs). Moreover, vehicles and several other infrastructures (e.g., weather stations, traffic control/help centers, etc.) need to communicate persistently for safety, surveillance, and cost optimization, requiring vehicle-to-everything (V2X) communication support including vehicle-to-vehicle , vehicle-to-infrastructure, and vehicle-to-pedestrian~\cite{dali_survey, li2018hybrid, peng2019vehicular}. 
Such connected-vehicle applications require scalable and long-range communication with wide availably of the communication medium. Additionally, connected-vehicle applications may require to transfer high-rate data for real-time decision making. TVWS spectrum may host a multitude of V2X applications given its spectrum availability, wider bandwidth, and long-rage communication support. Opportunities of vehicular networking including UAVs over TVWS has also been investigated by the works in~\cite{han2017vehicular, zhou2017tv, saleem2015integration}.}

\section{Research Challenges and Directions for TV White Space Networking}\label{sec:challenges}
In this section, we discuss the challenges for efficient and effective TVWS networking.

\paragraph{Interference Between Primary and Secondary Users}
While longer communication range eases the complexity of traditional wireless network topology and architecture, it also poses significant interference for the PUs, e.g, a SU operation in the TVWS may lead to complete shutdown of any nearby (in spectrum) licensed service.
Although, accessing the TVWS database or spectrum sensing are adopted by the SUs, it may not be enough while targeting applications that involve longer-range communication. 
Limiting the Tx power~\cite{FCC_first_order} of the SUs may lead to poor communication for demanding applications.
Hence, it demands the protocols be developed that can balance between Tx power and communication range, while completely avoiding interference between PU and SU.

\vspace{-0.3in} 
\revise{\paragraph{Coexistence of Homogeneous and Heterogeneous Applications}\label{sec:coexistence}
The FCC has envisioned that unlicensed operation in TVWS will surplus the development of protocols and applications currently available in unlicensed 2.4/5 GHz bands~\cite{FCC_first_order}.
Also, SNOW~\cite{snow_ton} has encouraged the research community to embrace TVWS for LPWANs, thus increasing the chances to enable IoT applications in a greater scale.
In future, several protocols may fight for coexistence in the TVWS spectrum. 
Coexistence issues may not only appear in heterogeneous networks but also within a network~\cite{li2016overhearing} or homogeneous networks~\cite{snow3, li2016multiuser, 7222470}.
Thus, it is important to develop protocols that will resolve coexistence issues in both homogeneous and heterogeneous network scenarios.}

\paragraph{Fragmentation of TV White Spaces}
Fragmentation poses significant challenge in adoption of TVWS protocols. Depending on bandwidth requirements, different applications will need a different number of TVWS channels.
Hence, both hardware and software protocols of these applications will need to use fragmented TVWS spectrum. 
Currently, very few protocols~\cite{snow_ton, wifi_like_connectivity,whitenet} aim to address that. However, they lack proper handling of fragmented TVWS.
While operating in the TVWS, protocols need to respect existing PUs and Sus operations in between the fragmented spectrum. 
Thus, protocols will require to adapt to variable Tx power in different parts of TVWS spectrum in the same application. This is a huge challenge for any protocol.
Also, traditional hardware limitations (sampling rate, antenna directions) may reduce adoption of fragmented TVWS spectrum, hence limiting the scalability of the applications due to cost and form factors.


\paragraph{Scarcity and Future Adaptation}
Compared to urban areas, rural and suburban areas tend have more available TVWS~\cite{wifi_like_connectivity}, making it challenging to continue uninterrupted operations in urban regions. 
Also in future, TVWS may be regulated to be used only by the PUs.
Such scenarios will make it impossible for applications to be hosted in the TVWS. Thus, protocols that wish to operate in the TVWS should be highly configurable and adaptable to relocate at any point of time on other unlicensed frequency bands, such as sub-1GHz (902 -- 928 MHz), 2.4/5 GHz. 
Currently, very few protocols can interchangeably operate on 2.4 and 5 GHz (e.g., Wi-Fi) band. It is a tremendous challenge for protocols/devices that are built only to operate in the TVWS to relocate. 
Thus, highly configurable software/hardware platforms for TVWS should gain considerable research focus.

\vspace{-0.3in} 
\revise{\paragraph{Antenna Design}
Due to the lower frequencies of TVWS, communication between two devices can span up to tens of kilometers~\cite{FCC_first_order}.
However, such benefit comes with a cost. From the physical form factor perspective, a device needs to mount a larger antenna to operate on lower frequencies such as TVWS. For efficient/correct reception, an antenna has to be on the order of $\ge\frac{1}{10}th$ of the wavelength of the transmitted signal ~\cite{antenna_book}. 
Wavelength (say, $\lambda$) of a signal is inversely proportional to the frequency of that signal~\cite{antenna_book}. 
This scenario becomes even worse when protocols require multiple antennas for desired QoS. Additionally, in multi-antenna protocols, antennas need to placed apart with a physical gap of $\ge\lambda/2$~\cite{snow2}. 
Additionally, it is infeasible for a small sensor device to mount larger and several antennas. Particularly, in vehicular networks, vehicles may not accommodate the above antenna separation and multi-antenna scenarios due to space limitation. 
Thus, research on smaller form-factor and energy efficient antennas are inevitable for enabling efficient TVWS networking.}

\vspace{-0.3in} 
\revise{\paragraph{Mobility}
Similarly, enabling mobility poses great challenges in the TVWS networking, especially in vehicular networks. Interestingly, none of the existing technique can be adopted directly to TVWS protocols since TVWS availability changes depending on time and geo-location~\cite{mobility1, mobility2}. 
Thus, a mobile device, which may be mounted on top of vehicles, operating in a TVWS channel may not be able to use the same channel in a different location. 
Additionally, FCC~\cite{FCC_first_order,fcc_second_order} mandates that each time a SU changes its location by 100 meters, it must contact the geo-location database or perform other techniques to determine the TVWS in its new location.
Enabling mobility in TVWS raises interesting and important challenges that need to be addressed properly to support applications such as V2X communications in the TVWS.}

\paragraph{Security}
Intrinsically, wireless communication is highly susceptible to security threats. The security requirements may differ from protocol to protocol depending on the communication range.
For example, the communication range of Bluetooth is between 5-10 meters, Near field communication (NFC) is few centimeters, making them naturally resistant to sniffing, spoofing, man-in-the-middle attacks. 
On the other hand, Wi-Fi and Zigbee incorporates several techniques~\cite{security_survey} to avoid security issues. 
TVWS protocols are currently being developed and lack proper security measures. Security threats become greater when the communication range increases~\cite{dali_survey}.  
Thus, research on security for protocols in the TVWS are significantly important as new threats will emerge due to longer communication range and possible isolation of devices in remote areas.

\section{Conclusion}\label{sec:conclusion}

\revise{The 2008 FCC ruling in the US on TVWS spectrum has opened up new opportunities for unlicensed operation in the TV band. In this paper, we have explored the properties of the TVWS spectrum including more availability, wider bandwidth, and excellent propagational characteristics that make TVWS suitable for longer-range, low-power, and large area applications such as sensing and monitoring applications, agricultural IoT applications, wireless broadband access, smart utility applications, location-based services, and transportation and logistics applications. To enable the above applications in the TVWS, we have identified several key research challenges that need to be explored and solved. Such challenges include handling interference between PU and SU, managing coexistence of homogeneous and heterogeneous applications, dealing with TVWS spectrum dynamics, TVWS antenna design, mobility, and security. In addition, we have provided a comprehensive review and a comparative study of existing TVWS networking protocols.}

\biboptions{sort&compress}
\bibliography{survey_bib}

\end{document}